\documentclass[10pt]{article}
\usepackage{amsmath,amssymb}
\usepackage[dvipdfmx]{graphicx}
\usepackage{bm}
\usepackage{color}
\usepackage{comment}
\usepackage{braket}

\def\mydate{27 June 2016}
\def\ignore#1{{}}

\newcommand{\beeq}{\begin{equation}}
\newcommand{\eneq}{\end{equation}}
\newcommand{\beqn}{\begin{eqnarray}}
\newcommand{\eeqn}{\end{eqnarray}}

\def\la{\raise.16ex\hbox{$\langle$}\lower.16ex\hbox{}  }
\def\ra{\raise.16ex\hbox{$\rangle$}\lower.16ex\hbox{} }
\def\go{\rightarrow}

\def\onehalf{ \hbox{$\frac{1}{2}$} }

\def\onethird{ \hbox{$\frac{1}{3}$} }
\def\twothird{ \hbox{$\frac{2}{3}$} }

\def\eff{{\rm eff}}

\def\SM{{\rm SM}}
\def\EM{{\rm EM}}
\def\diag{{\rm diag ~}}

\def\KK{{\rm KK}}

\def\vect{{\rm vec}}
\def\sp{{\rm sp}}

\def\psibar{ \psi \kern-.65em\raise.6em\hbox{$-$} }
\def\psibarl{ \psi \kern-.65em\raise.6em\hbox{$-$} \lower.6em\hbox{} }

\begin{document}

{\small \noindent \mydate    \hfill OU-HET 897}

\vskip 2.5cm

\baselineskip=15pt plus 1pt minus 1pt

\begin{center}
{\large \bf  Gauge-Higgs EW and Grand Unification\footnote{To appear in the Proceedings of 
{\it ``Conference on New Physics at the Large Hadron Collider''}, NTU, Singapore, 29 February - 4 March 2016.}}
\end{center}

\vskip 1.cm

\baselineskip=11pt plus 1pt minus 1pt

\begin{center}
{
Yutaka Hosotani
}

\vskip 5pt

{\small \it Department of Physics, Osaka University \\
Toyonaka, Osaka 560-0043, Japan} \\
{\small \it hosotani@phys.sci.osaka-u.ac.jp} \\

\end{center}



\baselineskip=10pt plus 1pt minus 1pt

\begin{abstract}
4D Higgs field is identified with the extra-dimensional component of 
gauge potentials in the gauge-Higgs unification scenario.
$SO(5) \times U(1)$  gauge-Higgs EW unification in the Randall-Sundrum warped space
is successful at low energies.  The Higgs field appears as an Aharonov-Bohm phase $\theta_H$
in the fifth dimension.  Its mass is generated at the quantum level and is finite.
The model yields almost the same phenomenology as the standard model
for $\theta_H < 0.1$, and predicts $Z'$ bosons around 6 - 10 TeV with very broad widths.  
The scenario is genelarized to $SO(11)$ gauge-Higgs grand unification.
Fermions are introduced in the spinor and vector representations of $SO(11)$.
Proton decay is naturally forbidden.
\end{abstract}




\baselineskip=13pt plus 1pt minus 1pt

\vskip 10pt
\leftline{\bf 1. Introduction}
\vskip 10pt

\noindent
We are looking for a principle for the 125 GeV Higgs boson which
regulates all Higgs couplings, explains electroweak (EW) symmetry breaking, and 
solves the gauge-hierarchy problem.  One possible answer is the gauge-Higgs 
unification.\cite{YH1,Davies1,Hatanaka1998}
One considers gauge theory in higher dimensions, say, in five dimensions.  
The 4D gauge fields, photon, $W$, and $Z$, appear as zero modes of the four-dimensional 
components of gauge potentials, whereas the 4D Higgs field is identified with the zero 
mode of the extra-dimensional component of gauge potentials.  
When the fifth dimension is compact and is not simply connected, 
the Higgs field appears as an Aharonov-Bohm (AB) phase, $\theta_H$,  
along the fifth dimension.

At the tree level the 4D Higgs field is massless.  At the quantum level the effective potential 
for  the AB phase, $V_\eff (\theta_H)$, becomes nontrivial, and a finite Higgs mass $m_H$ 
is generated.  At the same time dynamical breaking of the EW symmetry  takes place.
This is called the Hosotani mechanism.\cite{YH1}  The generated Higgs mass is finite, independent of
a cutoff scale and regularization method.  The gauge hierarchy problem is thus solved.

Construction of a concrete model of gauge-Higgs EW unification is highly nontrivial.
The standard model (SM) has $SU(2)_L \times U(1)_Y$ gauge symmetry, the Higgs
field is an $SU(2)_L$ doublet, and quarks and leptons are chiral in interactions.
All these features are naturally incorporated in the $SO(5) \times U(1)_X$  gauge-Higgs 
unification in the Randall-Sundrum warped 
space.\cite{ACP2005}-\cite{FHHOS2013}

\vskip 15pt
\leftline{\bf 2. $SO(5) \times U(1)_X$  gauge-Higgs EW unification}
\vskip 10pt

\noindent
Zero modes of the extra-dimensional component of gauge potentials must appear
as an $SU(2)_L$ doublet, and the custodial $SO(4)$ symmetry should result
in the Higgs part in four dimensions.  Further quark-lepton content must appear
chiral.  The minimal model is $SO(5) \times U(1)_X$ gauge theory defined on an
orbifold.  In five dimensions spacetime is either $M^4 \times (S^1/Z_2)$
or the Randall-Sundrum (RS) warped space.  Only in the RS space consistent
phenomenology is obtained.

The metric of the RS space is given by
\beeq
ds^2=  e^{- 2 \sigma(y)} \eta_{\mu\nu}dx^{\mu} dx^{\nu} + dy^2 
\label{metric1}
\eneq
where $\eta_{\mu\nu}=\mbox{diag}(-1,+1,+1,+1)$, 
$\sigma(y) = \sigma(y+ 2L)=\sigma(-y)$, and  $\sigma(y)=ky$ for $0 \le y \le L$.  
Topological structure of  the  RS space is the same as that of  $M^4 \times (S_1/ Z_2)$.
The points $y$, $-y$, and $y+2L$ are identified.  There appear two fixed points $(y_0,y_1)=(0,L)$.
The RS space is an AdS space sandwiched by the Planck brane (at $y=0$) and the TeV
brane (at $y=L$) with AdS curvature $\Lambda = - 6k^2$.

Only physical quantities need to be single-valued on orbifolds. $SO(5)$ gauge fields 
$A_M(x,y)$ obey 
\beqn
&&\hskip -1.cm
\begin{pmatrix} A_\mu \cr A_y \end{pmatrix} (x, y_j - y) =
P_j \begin{pmatrix} A_\mu \cr - A_y \end{pmatrix} (x, y_j + y) P_j^{-1} ~,  \cr
\noalign{\kern 5pt}
&&\hskip -1.cm
A_M(x, y+ 2L) = U A_M (x, y) U^{-1} ~, \quad U = P_1 P_0 ~.
\label{BCgauge1}
\eeqn
We take
\beeq
P_0 = P_1 = P_{\rm vec}=\diag (-1,-1,-1,-1, +1) ~.
\label{BC2}
\eneq
$U(1)_X$ gauge fields $B_M(x,y)$ satisfy a  condition similar to \eqref{BCgauge1}
where $P_j=1$.
At this stage $SO(5) \times U(1)_X$ symmetry breaks down to $SO(4) \times U(1)_X$.
Parity even-even modes (zero modes) appear in the $SO(4)$ block of $SO(5)$ $A_\mu$
and in $B_\mu$, and $SO(5)/SO(4)$ part of $A_y$ ($A_y^{j5}, j=1, \cdots, 4$). 
The latter is an $SO(4)$ vector,
or an $SU(2)_L$ doublet of $SO(4) \simeq SU(2)_L \times SU(2)_R$,  corresponding
to the 4D Higgs field in the SM.
In the gauge-Higgs unification the relevant quantity is the AB phase along the fifth dimension,
and is given by
\beeq
e^{i \hat \theta (x) / 2} = P \exp \bigg\{ ig_A \int_0^L dy \, A_y(x,y) \bigg\} .
\label{AB1}
\eneq

In the bulk ($0 < y < L$) we introduce quark and lepton multiplets $\Psi_a$ ($a=1, \cdots, 4$) 
in the vector representation of $SO(5)$ in each generation, and dark fermion multiplets 
$\Psi_{F_i}$ ($i=1, \cdots, n_F$) in the spinor representation.  They satisfy
\beqn
&&\hskip -1.cm
\Psi_a(x,y_j-y)=P_{\vect} \Gamma^5 \Psi_a(x,y_j+y),\cr 
\noalign{\kern 5pt}
&&\hskip -1.cm
\Psi_{F_i} (x, y_j -y)=
\eta_{F_i} (-1)^j P_{\sp} \Gamma^5\Psi_{F_i} (x, y_j + y),  
~~~ \eta_{F_i} = \pm 1 ,  
\label{BC3}    
\eeqn
where $P_{\sp} =\text{diag}  \, ( +1, +1, -1, -1 )$.
In addition, 
brane fermions $\hat \chi_\alpha$ in the $(\onehalf, 0)$ representation
and brane scalar $\hat \Phi$ in the $(0, \onehalf)$ representation 
of $SU(2)_L \times SU(2)_R$ are introduced on the Planck brane.
The brane scalar $\hat \Phi$ spontaneously breaks $SU(2)_R \times U(1)_X$
to $U(1)_Y$.  At the same time,  couplings on the Planck brane among
$\hat \Phi (x)$, $\hat \chi_\alpha$, and $\Psi_a(x,0)$ generate additional 
mass terms.  

The resultant symmetry is $SU(2)_L \times U(1)_Y$, which is subsequently 
broken to $U(1)_\EM$ by the Hosotani  mechanism
when $e^{i\hat \theta(x)}$ is not proportional to $I$.
Without loss of generality one may suppose that $\la A_y^{45} \ra \not=  0$, 
whereas $\la A_y^{j5} \ra = 0$  for $j=1,2,3$.
There appears one relevant AB phase $\theta_H$;
\beeq
\hat \theta_H(x) = \theta_H + \frac{H(x)}{f_H} ~,~~
f_H = \frac{2}{g_A} \sqrt{\frac{k}{z_L^2-1}}
=\frac{2}{g_w} \sqrt{\frac{k}{L(z_L^2-1)}} ~.
\label{AB2}
\eneq
Here $H(x)$ is the canonically normalized 4D neutral Higgs field and 
the warp factor  is  $z_L = e^{kL} \gg 1$.

\vskip 15pt
\leftline{\bf 3. Why gauge-Higgs unification?}
\vskip 10pt

\noindent
There are  good reasons for pursuing the gauge-Higgs unification.

\bigskip
\noindent
{\bf 
(a) Gauge principle governs the Higgs interactions.}

The 4D Higgs field is a part of the gauge potentials so that all Higgs interactions
emerges as gauge interactions in five dimensions.  Various couplings in four dimensions
appear as overlap integrals in the fifth dimension over three or more wave functions
of participating 4D fields, or as loop effects.   Yukawa couplings of quarks and leptons 
are in the former category, whereas the cubic and quartic couplings of Higgs self-interactions 
are in the latter category.

\bigskip
\noindent
{\bf 
(b) The finite Higgs boson mass $m_H$  is generated at the quantum level, free from a cutoff scale.
The gauge-hierarchy problem is solved.}

The Higgs field is a four-dimensional fluctuation mode of the AB phase $\theta_H$ in \eqref{AB1}
and \eqref{AB2}.  The effective potential $V_\eff (\theta_H)$ is flat at the tree level, 
but becomes nontrivial at the one loop level.  $\theta_H$-dependent part of $V_\eff (\theta_H)$
turns out finite, irrespective of the regularization method and cutoff scale.  This property is 
guaranteed by the gauge invariance in five dimensions.  When the global minimum of 
$V_\eff (\theta_H)$ is located at $\theta_H \not= 0$, the gauge symmetry is partially broken.
The whole scheme is called the Hosotani mechanism.\cite{YH1, YH2005, YH2006, HMTY2007}

The Higgs boson mass is given by
\beeq
m_H^2 = \frac{1}{f_H^2} \frac{d^2 V_\eff (\theta_H)}{d \theta_H^2} \bigg|_{\rm min},
\label{mH1}
\eneq
and is finite.  It is important to recognize that even though gauge theory in five dimensions
is not renormalizable, the $\theta_H$-dependent part of $V_\eff (\theta_H)$ is found to be
finite.  In the evaluation one has to take into account contributions from all KK modes.
The gauge hierarchy problem is solved.  This should be contrasted to the situation in 4D
gauge theory in which $m_H^2$ receives quantum corrections of $O(\Lambda^2)$ 
where $\Lambda$ is typically the GUT scale.

\bigskip
\noindent
{\bf 
(c) No vacuum instability.}

Scalar field theory in four dimensions is plagued by the vacuum instability problem
by quantum corrections.\cite{Degrassi2012}  
In the gauge-Higgs unification the effective potential for 
the Higgs field is given by $V_\eff [\hat \theta_H (x) ]$ where $\hat \theta_H (x)$ is 
given in \eqref{AB2}.  
The gauge invariance implies that $V_\eff (\theta_H + 2\pi) = V_\eff (\theta_H)$.
It follows that the global minimum is located somewhere in $0 \le \theta_H \le 2 \pi$,
or $0 \le H \le 2\pi f_H$, up to the periodicity.
There is no runaway instability.

\bigskip
\noindent
{\bf 
(d) Almost SM phenomenology at low energies.}

The $SO(5) \times U(1)_X$ gauge-HIggs unification in the RS space gives 
desired phenomenology at low energies and at the energy of 8 TeV LHC.
The SM matter content is reproduced at low energies.  Deviations of gauge
couplings of quarks, leptons,   $W$ and $Z$ from those in the SM turn out very 
tiny.\cite{HNU2010, HTU2011}

The Higgs couplings at the tree level receive corrections in an universal form.
All $HWW$, $HZZ$, $Hq \bar q$, $H\ell \bar \ell$ couplings are suppressed
by a common factor $\cos \theta_H$ compared to those in the SM.
For $\theta_H < 0.2$ the correction is less that 2\%, which is perfectly 
consistent with all data observed.\cite{HK2008}
 
One loop corrections also have been evaluated.  As explained in the next section,
the gauge-Higgs unification gives definitive predictions.  It will be found that 
the Higgs decay rates are almost the same as in the SM, as far as $\theta_H < 0.2$,
even when one loop corrections are included.\cite{FHHOS2013, FHH2015}

\bigskip
\noindent
{\bf 
(e) Dynamical EW symmetry breaking takes place.}

Once matter content is specified, $V_\eff (\theta_H)$ is unambiguously 
evaluated.  It is found that the minimal set of matter content described in Section 2 
leads to the dynamical breaking of the EW symmetry to $U(1)_\EM$.
In the RS space contributions of light quarks and leptons become negligible.
Contributions of gauge fields, top quark multiplet, and dark fermions are
relevant.  The existence of the top quark, whose mass is greater than $m_W$,
is crucial.\cite{FHHOS2013}

\bigskip
\noindent
{\bf 
(f) Fermion mass hierarchy is explained by $O(1)$ parameters in RS.}

There is a bonus in gauge-Higgs unification formulated in the RS space.
Large hierarchy in the fermion mass spectrum is explained in terms of
bulk mass parameters $c$ of fermion multiplets in the RS space.\cite{HTU2011} 
The third generation quark multiplet has $0< c < \onehalf$ so that the top 
quark acquires a mass, by the Hosotani mechanism, larger than $m_W$.
On the other hand, quark multiplets in the first and second generations and all lepton 
multiplets have $c > \onehalf $.  The mass hierarchy is easily generated with
a factor $z_L^{-c}$ for $c > \onehalf$.

\vskip 15pt
\leftline{\bf 4. Predictions}
\vskip 10pt

\noindent
At low energies it is hard to distinguish the $SO(5) \times U(1)_X$ gauge-Higgs unification 
scenario from the SM.   One needs to derive various predictions of
the gauge-Higgs unification which can be tested by experiments and observations.
The gauge-Higgs unification scenario is extremely restrictive, and therefore predictive.

\bigskip
\noindent
{\bf 
(a) Universality}

The model contains several parameters.  The metric of the RS space is specified with two 
parameters $k$ and $z_L=e^{kL}$, one of which is  fixed by $m_Z$.  
Two gauge couplings of $SO(5)$ and $U(1)_X$ are related to the $SU(2)_L$ coupling 
$g_w$ and $\sin^2 \theta_W$.
The bulk mass parameters of quark/lepton multiplets are determined by the quark/lepton spectrum,
in combination with brane interaction terms.  There are a few parameters in the dark fermion 
sector, one of them is fixed by the Higgs boson mass $m_H$.  In particular, the number of 
dark fermions, $n_F$, is arbitrary. In the minimal model, $z_L$ and $n_F$ may be treated
as free parameters.  Once $(z_L, n_F)$ is given, $V_\eff (\theta_H)$ is evaluated, from which
the location of its global minimum, the value of $\theta_H$, is determined.

One of the striking results in the $SO(5) \times U(1)_X$ gauge-Higgs unification is 
that many of the physical quantities depend, in a very good approximation, only on
$\theta_H$.  They are independent of the details in the dark fermion sector, particularly $n_F$.
This gives strong prediction power to the model.\cite{FHHOS2013}

First of all gauge couplings of the SM particles are almost the same as in the SM.
Deviations are typically less than 1\%, except for the couplings of $t$ and $b$ quarks.
The $\mu$-$e$ universality remains almost intact.  The deviation in the $WWZ$ coupling
is less than 0.1\%.  Three point Higgs couplings are given by \cite{HS2007, HK2008, Kurahashi2014}
\beeq
g_{HWW}, ~g_{HZZ}, ~g_{H q \bar q}, ~ g_{H \ell \bar \ell} \sim 
(\hbox{SM values}) \times \cos \theta_H ~.
\label{Hcoupling1}
\eneq

The KK mass scale, the masses of the first KK modes of $Z$ and $\gamma$, 
the mass of $SU(2)_R$ $Z_R$ are given by
\beqn
&&\hskip -1.cm
m_\KK \sim \frac{1352\,{\rm GeV}}{(\sin \theta_H)^{0.786}}  ~,  \cr
\noalign{\kern 7pt}
&&\hskip -1.cm
m_{Z^{(1)}} \sim \frac{1044\,{\rm GeV}}{(\sin \theta_H)^{0.808}}  ~,  \cr
\noalign{\kern 7pt}
&&\hskip -1.cm
m_{\gamma^{(1)}} \sim \frac{1056\,{\rm GeV}}{(\sin \theta_H)^{0.804}}  ~, \cr
\noalign{\kern 7pt}
&&\hskip -1.cm
m_{Z^{(1)}_R}\sim \frac{1038 \, \text{GeV}}{(\sin\theta_H)^{0.784}} ~.
\label{mass1}
\eeqn
The Higgs self-couplings arise at the one loop level.  The cubic and quartic couplings are
found to be
\beqn
&&\hskip -1.cm
\lambda^H_3 / {\rm GeV} \sim   26.7 \cos \theta_H  + 2.84  \cos^2 \theta_H  ~, \cr
\noalign{\kern 10pt}
&&\hskip -1.cm
\lambda^H_4 \sim  0.0214  + 0.0304 (\cos 2 \theta_H -1) + 0.00159 (\cos 4 \theta_H -1)~.
\label{HiggsCoupling}
\eeqn
These numbers should be compared with $\lambda_3^{H} = 31.5\,$GeV and 
$\lambda_4^H= 0.0320$ in the SM.  $\lambda^H_3$ vanishes at 
$\theta_H= \onehalf \pi$ due to the $H$ parity.\cite{HKT2009, HTU-Hparity}
The negative $\lambda^H_4$ for large $\theta_H$ does not imply the instability, as
$V_\eff (\theta_H)$ is bounded from below.

Once the value $\theta_H$ is determined, say, from the mass of $Z^{(1)}$, 
then all other quantities are predicted.

\bigskip
\noindent
{\bf 
(b) Loop corrections of KK modes in $H \go \gamma \gamma, gg, Z \gamma$ are finite and small.}

In the SM, the decay $H \go \gamma \gamma$ takes place through  one-loop processes
in which $W$ and top quark $t$  run.  In higher dimensional theory KK modes of $W$ and $t$
also run inside the loop.  Their contributions may add up to large, even diverging, corrections.
The same consern applies to $H \go g g$ and $H \go Z \gamma$.

In the gauge-Higgs unification miraculous cancellation takes place among contributions 
of KK modes.\cite{FHHOS2013, FHH2015}
The decay rate for $H \go \gamma \gamma$ is given by
\begin{align} 
	\Gamma ( H \to \gamma \gamma ) 
	&= \frac{\alpha^2g_w^2}{1024 \pi^3}\frac{m_H^3}{m_W^2} 
	\left|\mathcal{F}_W + \frac{4}{3} \mathcal{F}_t +n_F \mathcal{F}_F\right|^2, 
\label{2gamma1}
\end{align}
where $\mathcal{F}_W$, $\mathcal{F}_t$,  $\mathcal{F}_F$ represent contributions from
$W$, $t$, and dark fermion loops.  $\mathcal{F}_W$ is given by
\begin{align} 
\mathcal{F}_W &=\sum_{n=0}^{\infty}
\frac{g_{HW^{(n)} W^{(n)} }}{g_wm_W}\frac{m_W^2}{m_{W^{(n)}}^2} F_1(\tau_{W^{(n)}}) \cr
\noalign{\kern 3pt}	
&=\sum_{n=0}^{\infty}
	I_{W^{(n)}}\frac{m_W}{m_{W^{(n)}}} \cos\theta_HF_1(\tau_{W^{(n)}})
\label{2gamma2}
\end{align}
where $\tau_i = 4 m_i^2/m_H^2$ and 
$F_1(\tau) \sim 7$ for large $\tau$.    $I_{W^{(n)}}$ for large $n$ is approximately given by
\beeq
I_{W^{(n)}}\simeq (-1)^n \big\{0.0759 - 0.0065 \ln n + 0.0022 (\ln n)^2 \big\} ~.
\eneq
Similar behavior is found for $\mathcal{F}_t$ and $\mathcal{F}_F$ as well.
In other words the sum in each $\mathcal{F}$  behaves as
$\sum (-1)^n (\ln n)^\alpha /n$ ($\alpha =0,1,2$) and  rapidly converges.  
Moreover the contributions from $n \ge 1$ are suppressed by the ratio of
the electroweak scale to the KK scale.
The ratio of the amplitude to that with only zero modes is 
\begin{align}
	\frac{\mathcal{F}_W+\frac{4}{3}\mathcal{F}_t+4\mathcal{F}_F}
	{\mathcal{F}_{W^{(0)} \text{only}}+\frac{4}{3}\mathcal{F}_{t^{(0)} \text{only}}}=1.0027
\end{align}
at $\theta_H = 0.1153$. 
One finds that  the contributions of the KK modes are less than 1\% and negligible. 
For $H \go gg$ only the $t$ tower loops contribute, and the behavior is similar.

For $H \go Z \gamma$ the cancellation mechanism is more intricate.
In this case the KK number of particles running inside loops can change.
Miraculous cancellation occurs only when all possible diagrams are summed.
Numerically the correction due to KK modes amounts to only 0.07\% at $\theta_H = 0.1153$.

\bigskip
\noindent
{\bf 
(c) Signal strengths  in the Higgs decay}

As a consequence of \eqref{Hcoupling1} 
the decay widths of $H\to WW$, $H\to ZZ$, $H\to bb$ and $H\to \tau\tau$ are 
suppressed by $\cos^2\theta_H$ at the tree level.
The decay widths of the $H\to \gamma\gamma$, $H\to gg$ and $H\to Z\gamma$ are also 
suppressed by $\cos^2\theta_H$ with the cancellation mechanism among KK contributions 
taken into accout.  Consequently the branching ratios of the Higgs decay modes  
are almost the same as in the SM. 

The Higgs boson production is dominated by $gg\to H$, 
and the production cross section  is also suppressed by $\cos^2\theta_H$.
Therefore the signal strength of each decay mode $H \go j$, 
$\sigma(gg\to H)  B(H\to j)/ [\sigma(gg\to H)   B(H\to j) ]_\SM$, is 
approximately $\cos^2\theta_H$.
For $\theta_H \sim 0.1$, the deviation from the SM amounts to only 1\%.

\bigskip
\noindent
{\bf 
(d) $Z '$ bosons}

In the $SO(5)\times U(1)_X$ gauge-Higgs unification the first KK modes  
$Z_R^{(1)}$,  $Z^{(1)}$, and $\gamma^{(1)}$, appear as $Z'$ bosons in dilepton events at LHC.  
Here $Z_R$ is the neutral gauge boson associated with $SU(2)_R$, which does not have
a zero mode.   At LHC they are produced and detected as 
\[
 q \, \bar q \go Z_R^{(1)} , Z^{(1)} , \gamma^{(1)} \go e^+ e^- , ~ \mu^+ \mu^- ~.
\]
So far such events have not been observed, which put a constraint that their masses should be 
larger than  3$\,$TeV.

In the gauge-Higgs unification left-handed quarks and leptons are localized near the Planck brane,
whereas right-handed ones are localized near the TeV brane.  The first KK modes of gauge fields
are localized near the TeV brane so that right-handed quarks and leptons couple to 
$Z_R^{(1)}$,  $Z^{(1)}$, and $\gamma^{(1)}$ more strongly than quarks and leptons 
couple to $Z$ and $\gamma$.    For instance, couplings of right-handed $u$, $d$, and $e$
to $Z^{(1)}$ are about four times bigger than the corresponding couplings to $Z$.
All of $Z_R^{(1)}$,  $Z^{(1)}$, and $\gamma^{(1)}$ have
large widths.  $Z'$ events are not SM-like.
Masses and total decay widths of $Z_R^{(1)}$,  $Z^{(1)}$, and $\gamma^{(1)}$ are
summarized for $\theta_H=0.114$ and 0.073 in Table \ref{Z'table}.\cite{LHCsignal2014}

\begin{table}[htb]
\caption{Masses and total decay widths of  $Z^\prime$ bosons}
\begin{center}
\renewcommand{\arraystretch}{1.4}
{\begin{tabular}{|c|cc|cc|}
\hline
& \multicolumn{2}{|c|}{$\theta_H=0.114$}   & \multicolumn{2}{|c|}{$\theta_H=0.073$}    \\ 
\hline  
$Z'$            & $m$(TeV) & $\Gamma$(GeV) & $m$(TeV) & $\Gamma$(GeV)    \\ 
\hline
$Z_R^{(1)}$    & 5.73                & 482          & 8.00                & 553            \\ 
\hline
$Z^{(1)}$      & 6.07                & 342           & 8.61                & 494              \\ 
\hline
$\gamma^{(1)}$ & 6.08                & 886           & 8.61                & 1.04$\times10^3$             \\ 
\hline
\end{tabular} }
\end{center}
\label{Z'table}
\end{table}

\bigskip
\noindent
{\bf 
(e) Dark matter}

In this model  the lightest, neutral component  of $n_F$
$SO(5)$-spinor dark fermions  becomes  the dark matter of the  universe.\cite{DM2014}
The prediction concerning the dark matter, however, is not in the category 
of the universality explained above.  The prediction depends on the details
in the dark fermion sector.

The relic abundance of the dark matter determined by WMAP and Planck data
is reproduced,  below the bound placed by the direct detection experiment by LUX, 
by a model with one light and three heavier  ($n_F=4$)  dark fermions with the lightest one 
of a mass  from 2.3$\,$TeV  to 3.1$\,$TeV.
The corresponding  $\theta_H$  ranges from 0.097 to 0.074.

\vskip 15pt
\leftline{\bf 5. $SO(11)$ gauge-Higgs grand unification}
\vskip 10pt

\noindent
What is next? It is certainly necessary to incorporate strong interactions.  
The observed charge quantization in quarks and leptons,
for instance, is most naturally explained in the framework of grand unification.
It is desirable to have a unified theory of all gauge interactions.
There have been many attempts for gauge-Higgs grand unification, most of which
deals only with GUT symmetry 
breaking.\cite{Burdman2003}-\cite{Serra2011}

We would like to have a gauge-Higgs grand unification scenario which carries over good
features of $SO(5) \times U(1)$ gauge-Higgs EW unification.  $SU(6)$ theory, for instance,
does not meet this condition.  It does not give consistent EW phenomenology
at low energies.  We propose $SO(11)$ gauge-Higgs grand unification 
in the RS space.\cite{HY2015a}-\cite{FHY2016a}

\bigskip

\noindent
{\bf (a) Model}

$SO(11)$ gauge theory is defined in the RS space given by \eqref{metric1}.
$SO(11)$  orbifold boundary condition matrices $P_0$ and $P_1$ are given by
\beqn
&&\hskip -1.cm
P_0^{\rm vec}=\mbox{diag}(I_{10},-I_1) ~,~~ P_1^{\rm vec} =\mbox{diag}(I_4,-I_7)~,  \cr
\noalign{\kern 5pt}
&&\hskip -1.cm
P_0^{\rm sp} = I_{16} \otimes \sigma^3 ~, \hskip 1.2cm
P_1^{\rm sp} = I_2 \otimes \sigma^3 \otimes I_8 
\label{grandBCP1}
\eeqn
in vectorial and spinorial representations.  On the Planck brane $P_0$ breaks
$SO(11)$ to $SO(10)$, whereas on the TeV brane $P_1$ breaks $SO(11)$ to 
$SO(4) \times SO(7)$.  As a whole $SO(11)$ is broken to $SO(4) \times SO(6)$,
which is isomorphic to $SU(2)_L \times SU(2)_R \times SU(4)$.
$SO(11)$ gauge potentials satisfy \eqref{BCgauge1}.  
At this stage there appear parity even-even zero modes for $A_\mu$ in the
$SO(4) \times SO(6)$ block.  On the other hand zero modes of $A_y$
appear only for $A_y^{a \, 11}$ ($a=1 \sim 4$) components, which are
$SO(4)$ vector and $SO(6)$ singlet.   The zero modes of $A_y$ are identified with
the four-dimensional $SU(2)_L$ doublet Higgs field in the SM.

We introduce a brane scalar $\Phi_{\bf 16}$ on the Planck brane, in the spinor
representation of $SO(10)$. We suppose that $\Phi_{\bf 16}$ spontaneously
develops $\la \Phi_{\bf 16} \ra \not= 0$, which breaks $SO(10)$ to $SU(5)$
on the Planck brane.  As a consequence $SO(4) \times SO(6)$ symmetry is broken to
the SM symmetry, $G_\SM= SU(2)_L \times SU(3)_C \times U(1)_Y$.
$G_\SM$ is dynamically broken to $SU(3)_C \times U(1)_\EM$ by the Hosotani mechanism.

One immediate consequence is that all $SU(2)_L$, $U(1)_\EM$, $U(1)_Y$ charges,
and the Weinberg angle at the GUT scale are determined to be
\beeq
g_w = \frac{g}{\sqrt{L}} ~,~ e = \sqrt{\frac{3}{8}} \, g_w ~,~
g_Y = \sqrt{\frac{3}{5}}  \, g_w ~,~ \sin^2 \theta_W  = \frac{3}{8} ~.
\label{Wangle1}
\eneq

\bigskip

\noindent
{\bf (b) Fermions}

Fermions in the bulk are introduced in the spinor and vector representations of $SO(11)$.
In each generation of quarks/leptons $\Psi_{\bf 32}$, $\Psi_{\bf 11}$ and $\Psi_{\bf 11}'$
are introduced.  No additional brane fermions are necessary.  In a sense bulk and brane fermions
in the gauge-Higgs EW unification are unified in  grand unification.
The fermion fields obey
\beqn
&&\hskip -1.cm
\Psi_{\bf 32} (x,y_j - y)=  - \gamma^5 P_j^{\rm sp}  \Psi_{\bf 32} (x,y_j+y) ~, \cr
\noalign{\kern 10pt}
&&\hskip -1.cm
\Psi_{\bf 11} (x,y_j - y) = (-1)^j  \gamma^5 P_j^{\rm vec} \Psi_{\bf 11} (x,y_j+y) ~, \cr
\noalign{\kern 10pt}
&&\hskip -1.cm
\Psi_{\bf 11}^{\prime}  (x,y_j - y) 
=  (-1)^{j+1}  \gamma^5 P_j^{\rm vec} \Psi_{\bf 11}^{\prime} (x,y_j+y) ~.
\label{grandBCF1}
\eeqn

The content of these fermions is easily figured out.  One finds that for $\Psi_{\bf 32}$
\beqn
&&\hskip -1.cm
\Psi_{\bf 32} = \begin{pmatrix} \Psi_{\bf 16} \cr \Psi_{\overline{\bf 16}} \end{pmatrix} ,~
\Psi_{{\bf 16}} =    \begin{pmatrix}
\nu \cr e \cr \noalign{\kern 1pt} 
\hat e \cr \hat \nu \cr  \noalign{\kern 1pt}
u_k \cr d_k \cr \noalign{\kern 3pt}
\hat d_k \cr \hat u_k 
\end{pmatrix} , ~
\Psi_{\overline{\bf 16}} =    \begin{pmatrix}
\nu' \cr e' \cr \noalign{\kern 1pt} 
\hat e' \cr \hat \nu' \cr  \noalign{\kern 1pt}
u_k' \cr d_k' \cr \noalign{\kern 3pt}
\hat d_k' \cr \hat u_k' 
\end{pmatrix} , 
~(k=1 \sim 3) ,\cr
\noalign{\kern 5pt}
&&\hskip -1.cm
\hbox{zero modes :} \quad 
\begin{pmatrix}
\nu_{L} \cr e_{L} 
\end{pmatrix}, ~
\begin{pmatrix}
u_{kL} \cr d_{kL} 
\end{pmatrix} ,~ 
\begin{pmatrix}
\nu_{R}' \cr e_{R}'
\end{pmatrix}, ~
\begin{pmatrix}
u_{kR}' \cr d_{kR}'
\end{pmatrix} . 
\label{F32content}
\eeqn
Here the notation is such that $\hat e$, $\hat u$, and $\hat d$ have
charges $+1$, $- \twothird$, and $+ \onethird$, respectively.
For $\Psi_{\bf 11}$ and $\Psi_{\bf 11}'$
\beqn
&&\hskip -1.cm
\Psi_{\bf 11} = \begin{pmatrix}
\hat E ~ N \cr \hat N ~ E \cr \noalign{\kern 2pt}
D_k \, \hat D_k \cr \noalign{\kern 2pt}
S
\end{pmatrix}, ~
\Psi_{\bf 11}' =
\begin{pmatrix}
\hat E' ~ N' \cr \hat N' ~ E' \cr \noalign{\kern 2pt}
D_k' \, \hat D_k' \cr \noalign{\kern 2pt}
S'
\end{pmatrix}, \cr
\noalign{\kern 5pt}
&&\hskip -1.cm
\hbox{zero modes :} \quad 
D_{kR} ,  \hat D_{kR} , ~~
D_{kL}' ,  \hat D_{kL}' .
\label{F11content}
\eeqn
All quarks and leptons fit in the zero modes of $\Psi_{\bf 32}$.
Without the presence of $\Psi_{\bf 11}$ and $\Psi_{\bf 11}'$, however,
all quarks and leptons remain degenerate,  acquiring the same mass by the Hosotani mechanism.
To obtain the observed spectrum, one needs $\Psi_{\bf 11}$ and $\Psi_{\bf 11}'$.
An observed electron, for instance, is a linear combination of $e$, $e'$, $E$ and $E'$ fields.

\bigskip

\noindent
{\bf (c) Brane interactions and the fermion mass spectrum}

On the Planck brane $SO(10)$ gauge invariance is strictly maintained.
Bulk fermion fields which are parity even at $y=0$ can form scalar interactions
with $\Phi_{\bf 16}$.  $\Psi_{\bf 32}$ decomposes into $\Psi_{\bf 16}$ and $\Psi_{\overline{\bf 16}}$,
and $\Psi_{\bf 11}$ into $\Psi_{\bf 10}$ and $\Psi_{\bf 1}$ under $SO(10)$.
Participating fermion fields are $\Psi_{{\bf 16}L}$, $\Psi_{\overline{\bf 16}R}$,
$\Psi_{{\bf 10}R}$, $\Psi_{{\bf 1}L}$,  $\Psi_{{\bf 10}L}'$ and  $\Psi_{{\bf 1}R}'$.

Six types of brane interactions are allowed from the symmetry.
\beqn
&&\hskip -1.cm
S_{\rm brane} = \int d^5 x \sqrt{- \det G} \, \delta(y) \, \Big\{ 
- \kappa_1 \overline{\Psi}{}^{\prime}_{{\bf 1}R}  \,\Phi_{\bf 16}^\dagger  \, \Psi_{{\bf 16}L}
- \kappa_2
\overline{\Psi}{}_{{\bf 1}L}  \,\tilde \Phi_{\overline{\bf 16}}^\dagger  \, \Psi_{\overline{\bf 16}R}
\cr
\noalign{\kern 5pt}
&&\hskip 1.cm
- \kappa_3
 (\overline{\Psi}{}_{{\bf 10}R})_j   \, \hat{\tilde \Phi}_{\overline{\bf 16}}^\dagger  \,  
\Gamma^j \, \hat \Psi_{{\bf 16}L}
- \kappa_4
 (\overline{\Psi}{}^{\prime}_{{\bf 10}L} )_j   \, \hat\Phi_{\bf 16}^\dagger  \,  
\Gamma^j \, \hat \Psi_{\overline{\bf 16}R} \cr
\noalign{\kern 8pt}
&&\hskip 1.cm
- \mu_5 \overline{\Psi}{}^{\prime}_{{\bf 1}R}   \, \Psi_{{\bf 1}L}
-  \mu_6 \overline{\Psi}{}^{\prime}_{{\bf 10}L}   \, \Psi_{{\bf 10}R}
- ({\rm h.c.})  \Big\} ~.
\label{braneI1}
\eeqn
Here $\tilde \Phi_{\overline{\bf 16}} =\hat R \, \Phi_{\bf 16}^*$ transforms  as $\overline{\bf 16}$.
32 component notation has been adopted for $\hat \Psi_{{\bf 16}L}$, $\hat\Phi_{\bf 16}$ etc.
In general $\kappa_j$ and $\mu_j$ become 3-by-3 matrices in the generation space.
With $\la \Phi_{\bf 16} \ra \not= 0$,  \eqref{braneI1} generates six types of fermion mass terms,
in which $u$ and $u'$ do not appear.   The masses of up-type quarks are determined by
the bulk mass parameters $c_{\Psi_{\bf 32}}$ and $\theta_H$.  
It turns out that the $\kappa_2$ term is responsible for $m_\nu/m_e$, 
the $\kappa_3$ term for $m_e/m_u$, and the $\kappa_4$  and $\mu_6$ terms for $m_d/m_u$.
The observed fermion spectrum is reproduced by the Hosotani mechanism in combination
with the brane interactions.  Unfortunately there appear exotic light fermions associated with
$\hat u$, $\hat d$ and $\hat e$ in this scheme.

\vskip 15pt
\leftline{\bf 6. Forbidden proton decay}
\vskip 10pt

\noindent
Gauge-Higgs grand unification provides a new scheme of forbidding the proton 
decay.\cite{HY2015a, FHY2016a}
As seen in \eqref{F32content} and \eqref{F11content}, all quarks and leptons reside
in $\Psi_{\bf 32}, \Psi_{\bf 11}$ and $\Psi_{\bf 11}'$ as particles, but not as anti-particles.
In other words one can assign the $\Psi$-fermion number $N_\Psi$ such that all 
quarks and leptons have $N_\Psi=1$.  

The gauge interactions as well as the brane interactions \eqref{braneI1} preserve $N_\Psi$.
$N_\Psi$ is conserved.
The proton has $N_\Psi = 3$ whereas the positron has $N_\Psi = -1$ so that the 
process $p \go \pi^0 e^+$, for instance,  cannot take place.

This should be contrasted to 4D GUT.  In the four-dimensional $SO(10)$ GUT, for instance,
quarks and leptons are embedded in $\Psi_{{\bf 16}L}$.  In the notation in \eqref{F32content}, 
$\hat u_L$ is identified with $(u^c)_L$ or $(u_R)^c$ so that gauge interactions
do not conserve quark/lepton number, which induces the proton decay.
In the gauge-Higgs grand unification $u_L$ and $u_R$, for instance,  are embedded 
as zero modes of 5D fields $u$ and $u'$ in $\Psi_{\bf 32}$.  Both $u$ and $u'$ have $N_\Psi=1$.

\vskip 15pt
\leftline{\bf 7. Summary}
\vskip 10pt

\noindent
Gauge-Higgs unification is promising.  In the electroweak interactions we have 
$SO(5) \times U(1)_X$ gauge-Higgs EW unification.  It is consistent with 
data and observations at low  energies, including the data from 8$\,$TeV LHC.
It predicts $Z'$ bosons in the energy range 6 to 10$\,$TeV,
which should be observed at 14$\,$TeV LHC in a few years.
The deviation in the Higgs self-couplings from those in the SM is also predicted.

Grand unification is feasible in the gauge-Higgs unification scenario.
We have proposed $SO(11)$ gauge-Higgs grand unification, 
incorporating strong interactions. 
Comparison of symmetry structure in  the EW and grand unification is summarized in 
Table \ref{symmetrytable}.

The concrete model of gauge-Higgs grand unification with fermions
$\Psi_{\bf 32}, \Psi_{\bf 11}$ and $\Psi_{\bf 11}'$ reproduces the observed 
quark-lepton mass spectrum.  However, in the current minimal model there 
also appear light exotic fermions.   We need further elaboration of the model.

We add that there have been many advances in the gauge-Higgs 
unification.\cite{Kakizaki2014}-\cite{Hasegawa2016}
Dynamics of selecting orbifold boundary conditions has been explored.\cite{Yamamoto2014}
The Hosotani mechanism has been examined not only in the continuum theory, but also
on the lattice by nonperturbative simulations.\cite{Cossu2014, Forcrand2015, Knechtli2016}

\begin{table}[htb]
\caption{Symmetry structure in the gauge-Higgs EW and grand unification}
\begin{center}
\renewcommand{\arraystretch}{1.3}
{\begin{tabular}{ccc}
\hline
EW unification & & Grand unification      \\ 
\hline  
\noalign{\kern 5pt}
$SO(5) \times U(1)_X \times SU(3)_C$            & & $SO(11)$    \\ 
\noalign{\kern 5pt}
$\downarrow$  BC   &               & $\downarrow$  BC                \\ 
\noalign{\kern 5pt}
$SO(4) \times U(1)_X \times SU(3)_C$            & & $SO(4) \times SO(6)$    \\ 
\noalign{\kern 5pt}
 $\downarrow$ $\hat \Phi_{(0, \onehalf)}$     &         & $\downarrow$  $\Phi_{\bf 16}$     \\ 
\noalign{\kern 5pt}
~~ $SU(2)_L \times U(1)_Y \times SU(3)_C $ ~~  & ~   
& ~~ $SU(2)_L \times U(1)_Y \times SU(3)_C $  ~~    \\ 
\noalign{\kern 5pt}
$\downarrow$ $\theta_H$ &  &      $\downarrow$   $\theta_H$                    \\ 
\noalign{\kern 5pt}
$U(1)_\EM \times SU(3)_C $   &               &    $U(1)_\EM \times SU(3)_C $                \\ 
\noalign{\kern 5pt}
\hline
\end{tabular} }
\end{center}
\label{symmetrytable}
\end{table}

\vskip 10pt
\leftline{\bf Acknowledgments}
\vskip 10pt

\noindent
This work was supported in part  by Japan Society for the Promotion of Science, 
Grants-in-Aid for Scientific Research, No. 23104009 and No.\ 15K05052.

\def\jnl#1#2#3#4{{#1}{\bf #2},  #3 (#4)}

\def\Zphys{{\em Z.\ Phys.} }
\def\jssc{{\em J.\ Solid State Chem.\ }}
\def\jpsJ{{\em J.\ Phys.\ Soc.\ Japan }}
\def\ptps{{\em Prog.\ Theoret.\ Phys.\ Suppl.\ }}
\def\PTP{{\em Prog.\ Theoret.\ Phys.\  }}
\def\PTEP{{\em Prog.\ Theoret.\ Exp.\  Phys.\  }}
\def\JMP{{\em J. Math.\ Phys.} }
\def\NPB{{\em Nucl.\ Phys.} B}
\def\NP{{\em Nucl.\ Phys.} }
\def\PLB{{\it Phys.\ Lett.} B}
\def\PL{{\em Phys.\ Lett.} }
\def\PRL{\em Phys.\ Rev.\ Lett. }
\def\PRB{{\em Phys.\ Rev.} B}
\def\PRD{{\em Phys.\ Rev.} D}
\def\PRe{{\em Phys.\ Rep.} }
\def\AP{{\em Ann.\ Phys.\ (N.Y.)} }
\def\RMP{{\em Rev.\ Mod.\ Phys.} }
\def\ZPC{{\em Z.\ Phys.} C}
\def\SCI{\em Science}
\def\CMP{\em Comm.\ Math.\ Phys. }
\def\MPLA{{\em Mod.\ Phys.\ Lett.} A}
\def\IJMPA{{\em Int.\ J.\ Mod.\ Phys.} A}
\def\IJMPB{{\em Int.\ J.\ Mod.\ Phys.} B}
\def\EPJC{{\em Eur.\ Phys.\ J.} C}
\def\PR{{\em Phys.\ Rev.} }
\def\JHEP{{\em JHEP} }
\def\JCAP{{\em JCAP} }
\def\cmp{{\em Com.\ Math.\ Phys.}}
\def\JPA{{\em J.\  Phys.} A}
\def\JPG{{\em J.\  Phys.} G}
\def\NJP{{\em New.\ J.\  Phys.} }
\def\PoS{{\em PoS} }
\def\CQG{\em Class.\ Quant.\ Grav. }
\def\ATMP{{\em Adv.\ Theoret.\ Math.\ Phys.} }
\def\ibid{{\em ibid.} }

\renewenvironment{thebibliography}[1]
         {\begin{list}{\arabic{enumi}.$\,$}  
         {\usecounter{enumi}\setlength{\parsep}{0pt}
          \setlength{\itemsep}{0pt}  \renewcommand{\baselinestretch}{1.0}
          \settowidth
         {\labelwidth}{#1 ~ ~}\sloppy}}{\end{list}}

\def\reftitle#1{{\it #1, }}    

\end{document}